\documentclass[a4paper,10pt,twocolumn,preprint,3p]{elsarticle}
\usepackage{lineno,hyperref}
\modulolinenumbers[0]

 \journal{Preprint submitted to Nuclear Instruments \& Methods in Physics Research}

\begin{document}

\begin{frontmatter}

\title{Development of Spherical Proportional Counter for light WIMP search within NEWS-G collaboration}

%\author{Ali Dastgheibi-Fard}

\author[ifca]{Ali Dastgheibi-Fard \& Gilles Geribier \corref{thecorrespondingauthor}}
%\ead{ali.dastgheibi-fard@lpsc.in2p3.fr}  
\ead{ali.dastgheibi-fard@lpsc.in2p3.fr \& gilles.gerbier@queensu.ca}
%\author[queen]{Gilles Gerbier} 
\address[ifca]{LPSC site LSM, 1125 Route de Bardonneche; 73500 Modane-France}

\cortext[thecorrespondingauthor]{Corresponding author. Tel.: +33 479 05 5455}
\author{on behalf of the NEWS-G Collaboration}

\author[mysecondaryaddress]{}
\cortext[mycorrespondingauthor]{Ali Dastgheibi-Fard}
%\ead{ali.dastgheibi-fard@lpsc.in2p3.fr \& gilles.gerbier@queensu.ca}

\begin{abstract}
The Spherical gaseous detector (or Spherical Proportional Counter, SPC) has a broad range of applications.  In this work, we will focus on the light WIMP Dark Matter particle search, that is, below a few GeV. The NEWS-G collaboration operates a 60 cm diameter detector at Modane underground laboratory (LSM). Running the detector at 3 bars of  Neon gas, some competitive limits  have been set  down to 0.5 GeV WIMP mass  thanks to an energy threshold of around 100 eV. To reach better performances, the next generation of SPC (140cm), NEWS-G\_SNO, is under construction. We will focus here on the fabrication method, the different low energy calibration methods using  $^{37}$Ar, neutron and UV laser,  on the cleaning methods to remove the surface contamination and mitigation of $^{210}$Pb bulk contamination.
%\LaTeX\ manuscript.
\end{abstract}

\begin{keyword}
Dark Matter search, Light WIMP search, Spherical Proportional Counter, Spherical detector
%\texttt{elsarticle.cls}\sep \LaTeX\sep Elsevier \sep template
\MSC[2019] 00-01\sep  99-00
\end{keyword}

\end{frontmatter}

%\linenumbers

\section{Introduction}

%\paragraph{} 

Dark Matter (DM) is currently seen as an unavoidable piece of the Universe puzzle and is central to a lot of new theories and models of particle physics  \cite{WimpJ-L.Feng}.

The WIMP (Weakly Interacting Massive Particles) Dark Matter is one of the most probable candidate and ''WANTED'' particle by the DM community. So, the search for  WIMP's  is under intense development and relies on the detection of low energy recoils (keV scale)  produced by the elastic interaction of WIMPS with the nuclei of the detector. Although the whole possible DM mass range covers 90 order of magnitude, the main focus has been put on  the WIMP mass range  between few tens and few hundreds of GeV. To cover a large range of masses,  many experiments address the search of WIMPs from a few GeV down to few 100 MeV. The presently studied detector, a Spherical Proportional Counter (SPC) -gaseous detector- , initially proposed by I.Giomataris \cite{giomatarisNovel2008}, will allow to explore such new parameter space for Dark Matter. \ 

NEWS-G (New Experiments With Spheres - Gas)  \cite{NEWSgHomepage} is a direct DM detection experiment using SPCs  to find a signature of DM at the GeV scale.  Since low-background, low-energy threshold are keys for their detection, the NEWS-G detectors are fabricated with very radiopure material - mainly copper - and operated  at energy threshold below down to single electron detection.\

NEWS-G\_LSM, a 60 cm diameter SPC installed at Laboratoire Souterrain de Modane (LSM) \cite{LsmWebsite} is under operation since 2012. Thanks to a very low background and an energy threshold of around 100 eV, a 41 days physic run  with gas mixture based on Ne and 0.7 \% of CH4 at 3 bars allowed to produce  a competitive limit on light WIMP searches  \cite{QuentinFirstResultDM}. 

A larger and lower activity detector, a 140 cm SPC, NEWS-G\_SNO , is being built together with improved compact shielding and will be installed in SNOLAB.  
Since the fabrication of sphere and lead shield took place in France and to insure that all manufacturing construction parameters have been taken into account,  the preliminary test and calibration will be completed at LSM in next months of 2019, and the setup  will be operated in late 2019 at SNOLAB \cite{SNOLABWebsite}.\

Other applications of SPC's range from radon emanation gas monitoring, neutron flux, gamma counting and spectroscopy, to neutrino physics, double beta decay \cite{SPCDBD} and coherent neutrino scattering. 
Expertise in these various applications is shared within the NEWS-G  collaboration.

\section{Detector description}
%\paragraph{}

The SPC consists of a grounded spherical  metallic vessel (15 cm to 140 cm in diameter) , the cathode, and a small ball (1 mm to 16 mm in diameter), the anode, placed at the centre of the vessel at the end of a grounded metallic rod. A wire connects the anode to the high voltage  and the  signal is read-out through a capacitor. The electric field varies as 1/r$^2$ inside  the spherical vessel, where " r " is  radius. To compensate  the inhomogeneous electric field close to the rod, a second electrode is designed upstream to the ball. More details of recent development and application are given in Ref. \cite{Arnaud2016novel}.\

\begin{figure}[htbp]
\begin{center}

    \begin{tabular}{cc}
     
      \includegraphics[width=5cm]{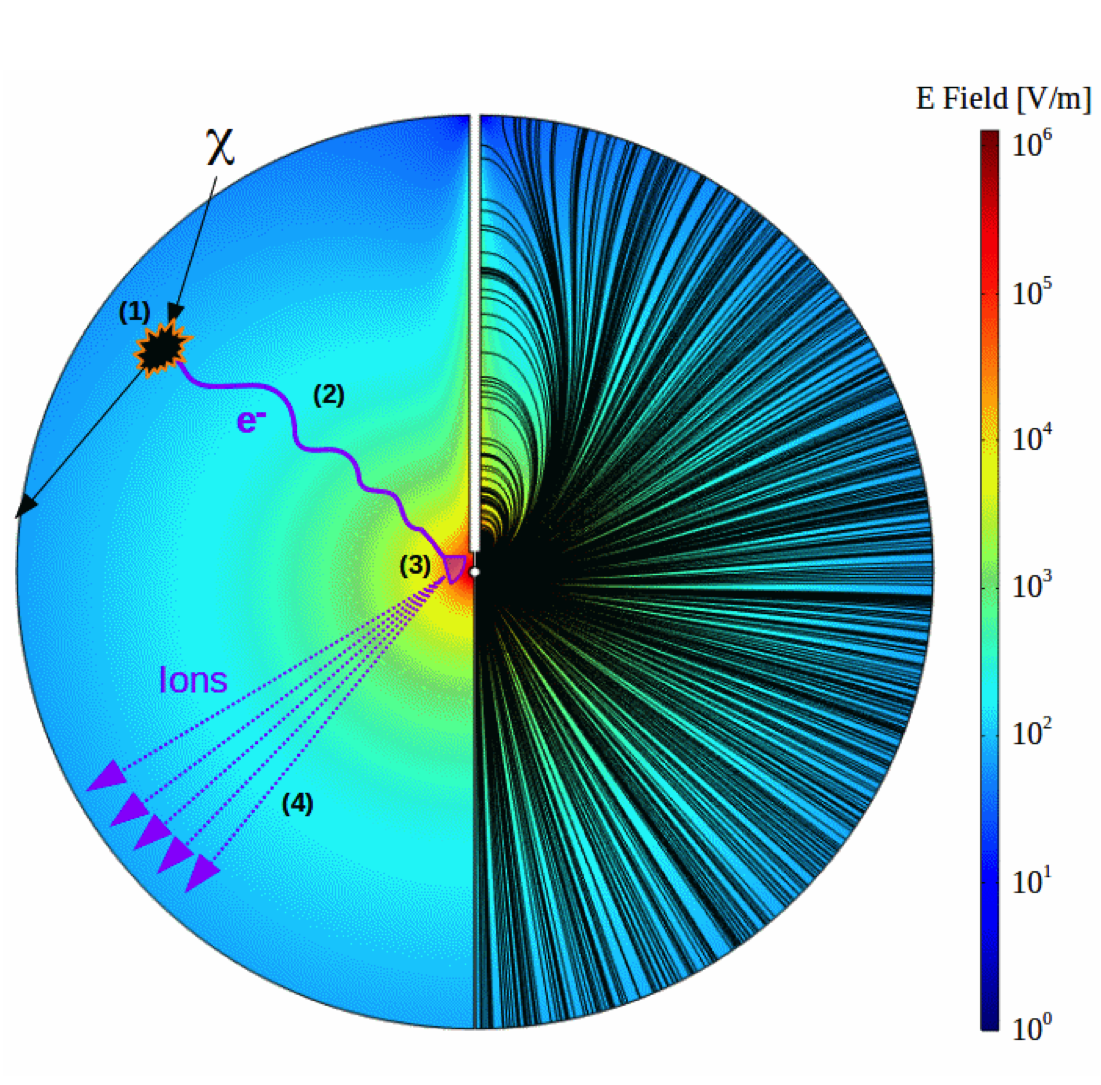}
      
         \end{tabular}
\caption{\textit{Principle of operation of spherical gas detector }}
\label{Fig:DetectorAndElecField}
\end{center}
\end{figure}

The electrons created in the gas volume drift to the central anode. A few millimetres from the ball, in region of high field, an avalanche occurs and  induces the signal (figure \ref{Fig:DetectorAndElecField}). 

\begin{figure}[htbp]
\begin{center}

    \begin{tabular}{cc}
     
      \includegraphics[width=7cm]{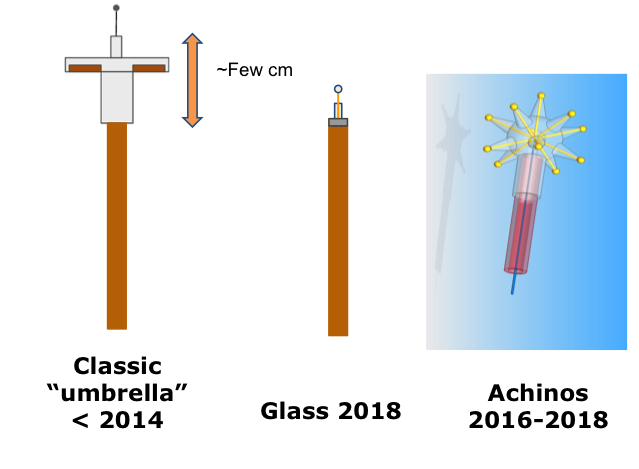}
      
         \end{tabular}
\caption{\textit{Different sensor developments }}
\label{Fig:Sensors}
\end{center}
\end{figure}

\underline {\textbf{Sensor:}} The sensor, main part of the detector, consists of a ball connected to the HV  and a second electrode (umbrella) which aims at  keeping an homogenous 1/r$^2$ electric field . In figure \ref{Fig:Sensors}, the evolution of the various sensors fabricated and used in SPC's. Recently, a multi-ball sensor is under test for new  generation of SPC \cite{Achinos}.

The main features of the SPC include a very low capacitance, a low energy threshold independent of the volume (down to single ionization electron), a good energy resolution, robustness and a single detection readout channel. 

\section{Material selection}
%\paragraph{}

Copper is the main component of the spherical vessel in the SPCs; it represents  more than 90 \% of the weight and 95 \% of the detector surface. This is the reason why all of our efforts go to choose  ultra radiopure copper, to store it in the underground laboratory between steps of fabrication,  to take care of fabrication method and to perform chemical cleaning  to reduce all background sources coming from copper surfaces. \\

\begin{figure}[htbp]
\begin{center}

    \begin{tabular}{cc}
     
      \includegraphics[width=7cm]{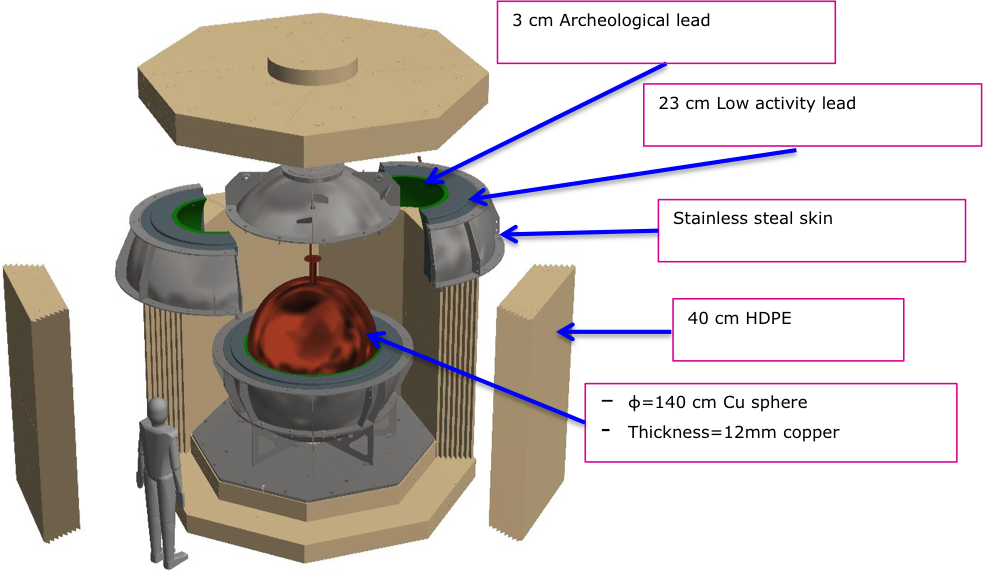}
      
         \end{tabular}
\caption{\textit{ NEWS-G\_SNO vessel and shielding compositions}}
\label{Fig:NEWSgSNO}
\end{center}
\end{figure}

\underline {\textbf{Fabrication:}} The NEW-G\_LSM has been fabricated with NOSV copper, all measurements are referenced in  \cite{TheseAli}. Concerning the 140 cm diameter, used copper is  the commercial Aurubis™ Oxygen Free Copper C10100 . For both detectors, the hemispheres were welded with electron beam not to add  extra material. The NEWS-G\_SNO will use a compact shield as shown in figure \ref{Fig:NEWSgSNO}.\\

The ICPMS measurement at PNNL \cite{SiteWebPNNL} allowed to determine the activity of Uranium and Thorium in copper:\

$^{238}$U = 1 - 5 $\mu Bq/kg$\

$^{232}$Th = 7 - 25 $\mu Bq/kg$\\

\underline {\textbf{Cleaning:}} Although the copper was chosen  with a very pure quality,  it still had a non negligeable  $^{210}$Pb bulk contamination. The activity of  $^{210}$Pb has been  measured to be  around 40 mBq/kg, thanks to an agreement with the XMASS group \cite{XMASS} who developed  an alpha spectrometry method using an XIA detector. Through simulations, the measured bulk  contaminations has been anticipated to be responsible for  80 \% of the background in the region of Interest for light DM search, below a few keV.\

For NEWS-G\_LSM,  chemical cleaning using nitric acid solution based (resp. 17\% and 30\%) allowed to reduce the surface contamination by two order of magnitudes \cite {AliTipp}.\

For the NEWS-G\_SNO detector vessel, to mitigate impact of contaminations of $^{210}$Pb,  $^{210}$Bi beta’s and Xrays, a 500 $\mu m$ pure copper was deposited by electroplating on the inner surface of the two hemispheres before welding. Electropolishing was also performed at beginning to remove the surface layer.

\section{Calibrations and results}
%\paragraph{}

For the SPC energy calibrations,  three calibrations methods were developed. 

\begin{figure}[htbp]
\begin{center}

    \begin{tabular}{cc}
     
      \includegraphics[width=5cm]{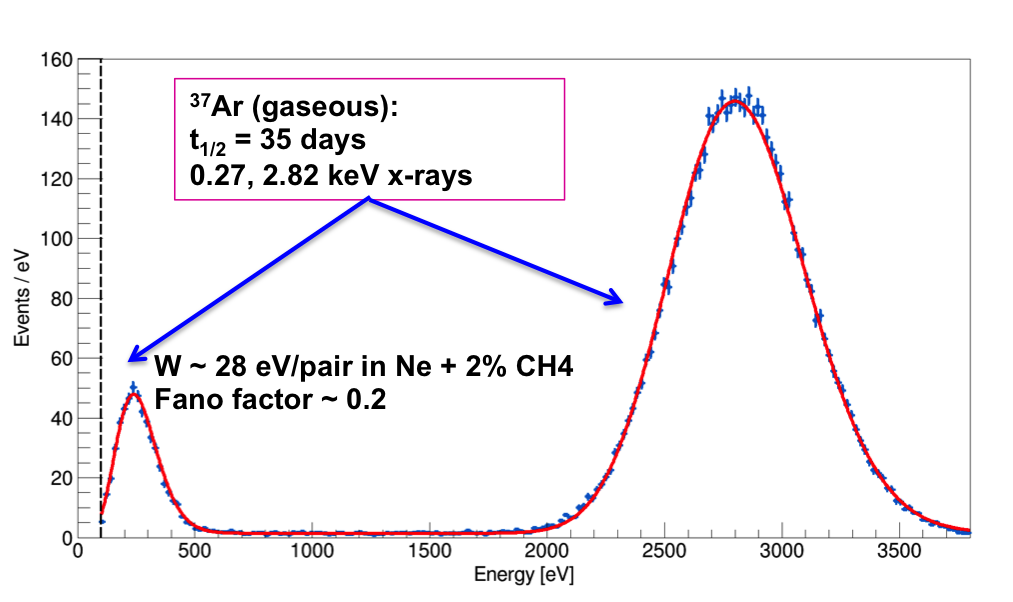}
      
         \end{tabular}
\caption{\textit{$^{37}$Ar calibration}}
\label{Fig:Ar37}
\end{center}
\end{figure}

The volume energy calibration is performed with low energy X rays from $^{37}$Ar gas, made from fast neutron irradiation of $^{40}$Ca in the slowpoke reactor of RMCC \cite{RMCC}. The detector filled with the appropirate target gas then $^{37}$Ar gas, which fills all volume and allows to measure the homogeneity of response of the detector. $^{37}$Ar decays through electron capture with t$_{1/2}$ = 35 day, giving $^{37}$Cl, emitting  single X rays X$_{k \alpha 1}$ of 2.6 keV and X$_{L \beta 3}$ of 260 eV (see  figure \ref{Fig:Ar37}) .

\begin{figure}[htbp]
\begin{center}

    \begin{tabular}{cc}
    
      \includegraphics[width=5cm]{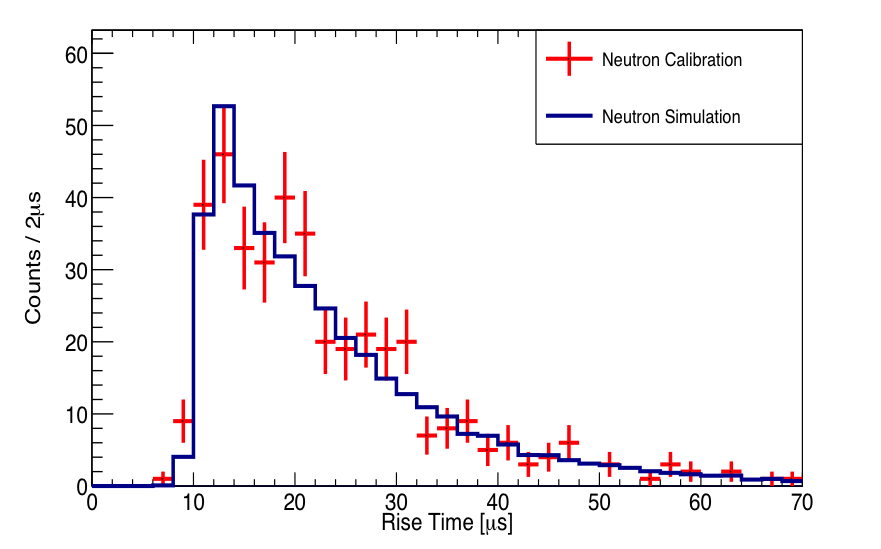}
     
         \end{tabular}
\caption{\textit{rise time distribution of events in the energy range [150, 250] eVee for events from neutron calibration}}
\label{Fig:NeutronCalib}
\end{center}
\end{figure}

A neutron source has  been used to calibrate the risetime distribution of low energy events, reflecting the radius distribution of point like energy deposition  inside the sphere. Figure \ref{Fig:NeutronCalib} shows the rise time spectrum at the lowest [150, 250] eVee energy bin used to set limits on  sub- GeV/c$^{2}$ WIMPs  together with the quite good agreement with our modeling of detector response. 

Also, thanks to the use of a pulsed 213nm Laser, we developed an original chracteriszation  methodology allowing to measure simultaneously the single electron response, the first ionization potential "w", the gain of the detector, the $\theta$ parameter of Polya distribution $"\theta Polya"$ and to monitor gain stability within 1 \%.\cite{LaserPaper}

\underline {\textbf{Dark Matter search:}} The result of a Dark Matter search run taken with NEWS-G\_LSM (60 cm sphere) and the analysis strategy is presented in this reference \cite {QuentinFirstResultDM}.\

\begin{figure}[htbp]
\begin{center}

    \begin{tabular}{cc}
     
      \includegraphics[width=8cm]{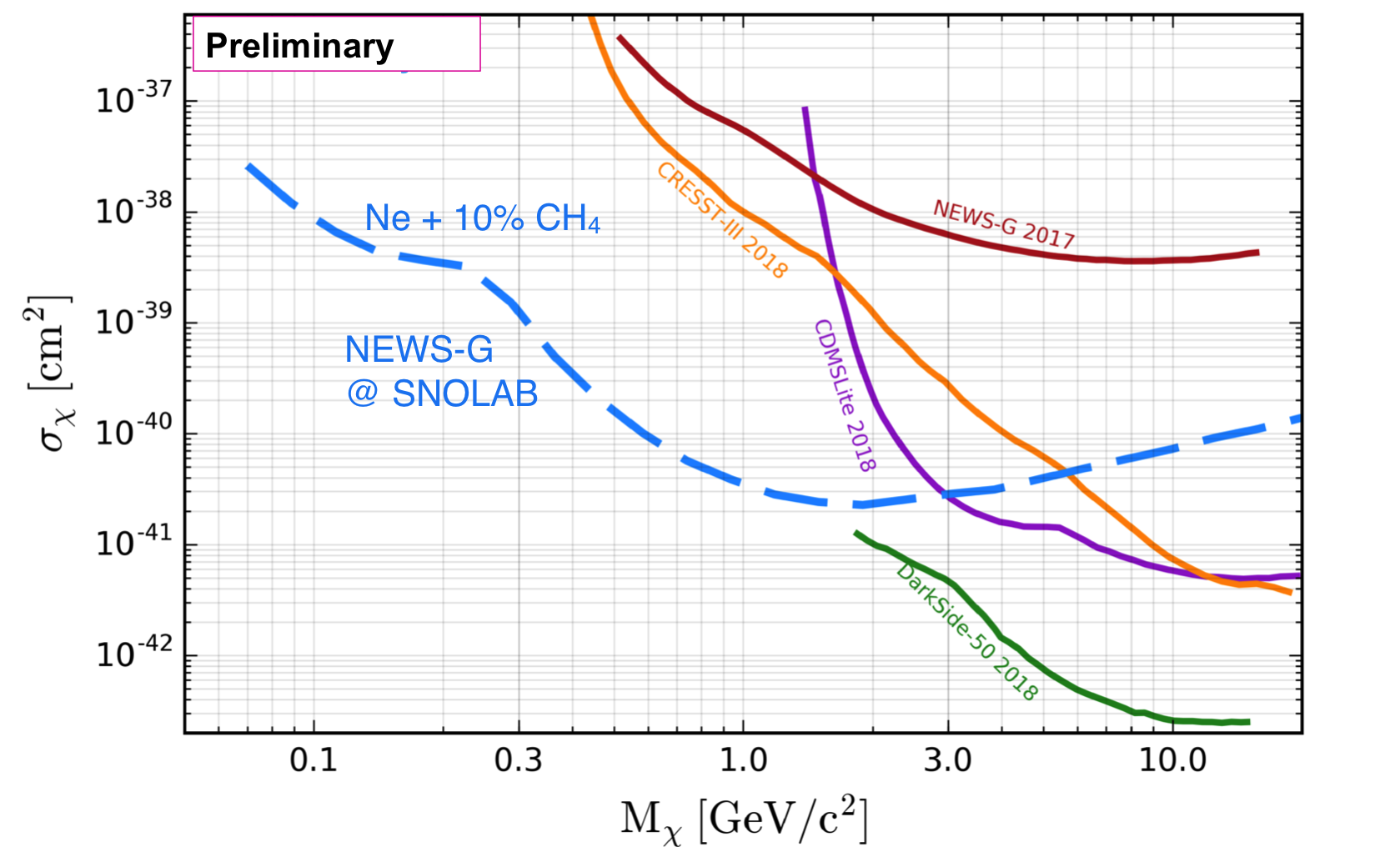}
  
         \end{tabular}
\caption{\textit{Result of NEWS-G\_LSM and projection of NEWS-G\_SNO sensitivity for WIMP search using Ne + (10\%) CH4 gas mixture }}
\label{Fig:DMProgection}
\end{center}
\end{figure}

Figure \ref{Fig:DMProgection} summarizes the already obtained result in the cross section-WIMP mass parameter space together  with the projection of the expected results with NEWS-G\_SNO, assuming an exposure  of 20 kg.d with Ne + 10\% CH4 gas mixture,  a single electron energy threshold and  the anticipated  background budget based on simulations.\\

{\bf{Acknowledgements} }

We  wish to thank the help of Saclay and Laboratoire Sourterrain de Modane teams of NEWS-G collaboration. The NEWS-G\_LSM has been partially funded within the European Commission astroparticle program ILIAS (Contract R113-CT-2004- 506222).\\

\bibliography{mybibfile}

\end{document}